# 1D/2D hybrid Te/Graphene and Te/MoS$_2$: multifaceted broadband photonics and green energy applications

Tuhin Kumar Maji[1], Kumar Vaibhav[2], Anna Delin[3], Olle Eriksson[4, 5]*, Debjani Karmakar[4,6]*

[1] Department of Physics, Indian Institute of Science Bangalore, Bangalore, 560012, India
[2] Computer Division, Bhabha Atomic Research Centre, Trombay, Mumbai 400085, India
[3] Swedish e-Science Research Center (SeRC), KTH Royal Institute of Technology, SE-10044 Stockholm, Sweden
[4] Department of Physics and Astronomy, Uppsala University, Box 516, SE-75120 Uppsala, Sweden
[5] School of Science and Technology, Örebro University, Fakultetsgatan 1, SE-70281 Örebro, Sweden
[6] Technical Physics Division, Bhabha Atomic Research Centre, Trombay, Mumbai 400085, India

**Corresponding Author:** debjan@physics.uu.se, olle.eriksson@physics.uu.se

**Abstract:** In this letter, we highlight the enhanced functionalization of the electronic and optical properties in the hybrid heterojunction of 1D Tellurene with 2D monolayer of Graphene and $MoS_2$ in both lateral and vertical geometries, having potential applications in the field of photonics and energy harvesting. The structural geometries of the lateral and vertical assemblies are optimized with a comparative and systematic analysis of the energetics of the different positional placement of the 1D system with respect to the hexagonal 2D layer of the substrate. The 1D/2D coupling of the electronic structure in this unique assembly enables the realization of the four different types of heterojunctions, *viz.* type I, type II, *Z*-scheme and Schottky type, with the band-alignments being entirely dependent upon the stacking geometry of 1D Tellurene with respect to the 2D monolayer. With the static and time-dependent first-principles calculations, we indicate the potential applications of these hybrid systems in broadband photo detection and absorption, covering a wide range of Infra-red to visible (IR-Vis) spectrum from 400 to 2500 nm. With an inherent effective separation and migration of photo-generated charge carriers, this 1D/2D assembly can additionally be utilized in green energy harvesting.

The genesis of van der Waal heterojunctions with systems of different dimensionality offers a possibility of flexible atomic scale integration of materials with widely distinct electronic properties and thus widens the horizon of diverse photonics and green-energy applications [1-3]. Befitting these attributes, 1D/2D hybrids offer the advantages of integrating the leverages of 1D nanostructures like better electron transport, higher aspect ratio and larger specific surface area for an improved absorption of light with the endowed plus-points of 2D materials like larger concentration of active sites, shorter diffusion length and high mobility. Thus, such systems expedite a faster transfer of free charge, facilitating the separation and migration of photo-generated carriers [4-8]. The recent surge of multifarious applications of hybrid materials in the field of photo-detection and energy harvesting comprises both the lateral and vertical stacking geometries, synthesized by ex- and in-situ modes [9-12]. Amongst the numerous reported 1D/2D hybrids, one should mention the significant applications like the enhanced photo-detectivity of Te-nanorod/Graphene, Se/InSe [1,13], improved optoelectronic activity of 1D double perovskite/2D materials [14], utilization of $WO_{3-x}$/Graphene as an pressure sensor [15], Carbon nanotube/ Graphene or $MoS_2$ as a thermal sensor [16], $Bi_2S_3/MoS_2$ for photodetector and rectifier [17,18], $TiO_2/Bi_2WO_6$ for gas-sensor [19] and $C_3N_4$/Graphene for improved anode-material for Li-ion batteries [20,21]. Moreover, 1D/2D hybrids are also well known to have widespread applications in the field of green and sustainable energy harvesting by providing better pathways for the separation and migration of carriers for systems like $TiO_2/ZnIn_2S_4$ [22], $CdS/ZnIn_2S_4$ [23], $CdS/MoS_2$ [11], TiS/Mxene [24], $FeSe_2/MoSe_2$ [25], carbon nanotune/phosphorene [26] and many more [27-30].

The photoactive performances of semiconducting photodetectors highly depend on the optimized regime of the wavelength detection, specific to a particular combination of the hybrids. While there are numerous systems like silicon, InGaAs or GaN, having utilizations in the ultraviolet (UV) and visible (Vis) range, materials encompassing the full Infra-red (IR) range are less frequent. IR detectors are categorized as photonic and thermal detectors with the near (NIR) plus mid (MIR) and the far-IR(FIR) ranges respectively [31,32]. The faster performance and compact sizes of the photonic IR detectors has stimulated the search for effective materials, possessing optical activities in the full IR-range. Suitable materials may reinstate the costly and complicated synthesis of commonly used semiconductor alloy materials like HgCdTe or semiconductors like Pb chalcogenides or InSb [33-37] with simple chemical synthesis.

After the advent of novel 2D materials like graphene or transition metal dichalcogenides (TMDC), broadband photo-detection has gained a new pace [38,39]. Graphene is well-known for its properties of ultra-broadband detection as a result of its zero band-gap nature [40]. However, the short lifetime of the hot carriers as a result of their ultrafast recombination and the resulting limited responsivity of metal-contacted bare graphene detectors and *p-n* junctions [41-44], there are continuous efforts to enhance the performance upon coupling with microcavities [45,46], plasmonic structures [47,48], quantum dots [49] or fabricating stacked structures with TMDC [50,51] by increasing the lifetime of the carriers through physical separation.

In this letter, we indicate the exorbitant application potential of the hetero-junction (HJ) of 1D single helix of Tellurene, the building-block of Te-nanorod (Te-NR), with the 2D monolayer of graphene (G) or $MoS_2$ (M) in lateral and vertical stacking geometries and analyze their comprehensive electronic structure and optical responses. Four different types of 1D/2D HJs are explored, *viz.*, 1) Te-NR on Graphene in lateral (Te/G(L)) and 2) in vertical geometry (Te/G(V)), 3) Te-NR on $MoS_2$ in lateral (Te/M(L)) and 4) in vertical geometry (Te/M(L)).

α-Te is an extremely narrow band gap semiconductor in its bulk form having a band gap of ~ 0.35 eV in the FIR range [52-54], as shown in Fig S1(a). [54] The bulk structure of α-tellurene consists of a van der Waal bonded hexagonal array of covalently bonded c-axis oriented single helix of Te in the a-b plane [54]. An isolated single helix or Te-NR possesses a band gap of ~ 1.2 eV (Fig S1(b)). While graphene is a well-accomplished zero band gap system, the monolayer(ML) of $MoS_2$ with a band gap of ~1.8 eV is at the boundary of NIR and Vis wavelengths (Fig S1). The 1D/2D HJ in the lateral and the vertical stacking geometries are constructed from the Te-NR and the 8×4×1 supercell of ML of graphene and $MoS_2$ after keeping an interfacial strain, which is < 5%, as shown in Fig S2. [54] The energy optimization procedure corresponding to the construction of the HJs are summarized in Fig 1, where the formation energy of the hybrid system at different position are calculated by using the relation $E_{form} = E_{1D/2D} - E_{1D} - E_{2D}$. The lowest energy configurations for the vertical placements are ascertained in two steps. First, the energetics corresponding to the placement of a single Te-NR is assessed with respect to the G and M hexagons. Next, for higher densities of Te-NRs, the optimal distance between the Te-NRs are calculated for an assembly of 8×4×1 monolayer of G and M and four NRs. The placement of the Te-NRs is done in such a way so as to avoid the electronic overlap of the periodic images from the consecutive cells. Fig 1(a) and (b) presents a comparison of the formation energies of the Te/G(V) and Te/M(V) HJs for the placement of the Te-NR with respect to a single hexagon. The pink arrow at the bottom designates the relative direction of the displacement of the Te-NR, starting from one vertex of the hexagon to the diametrically opposite one. The top view of the NR is a red equilateral triangle and its end point is marked in green as E.

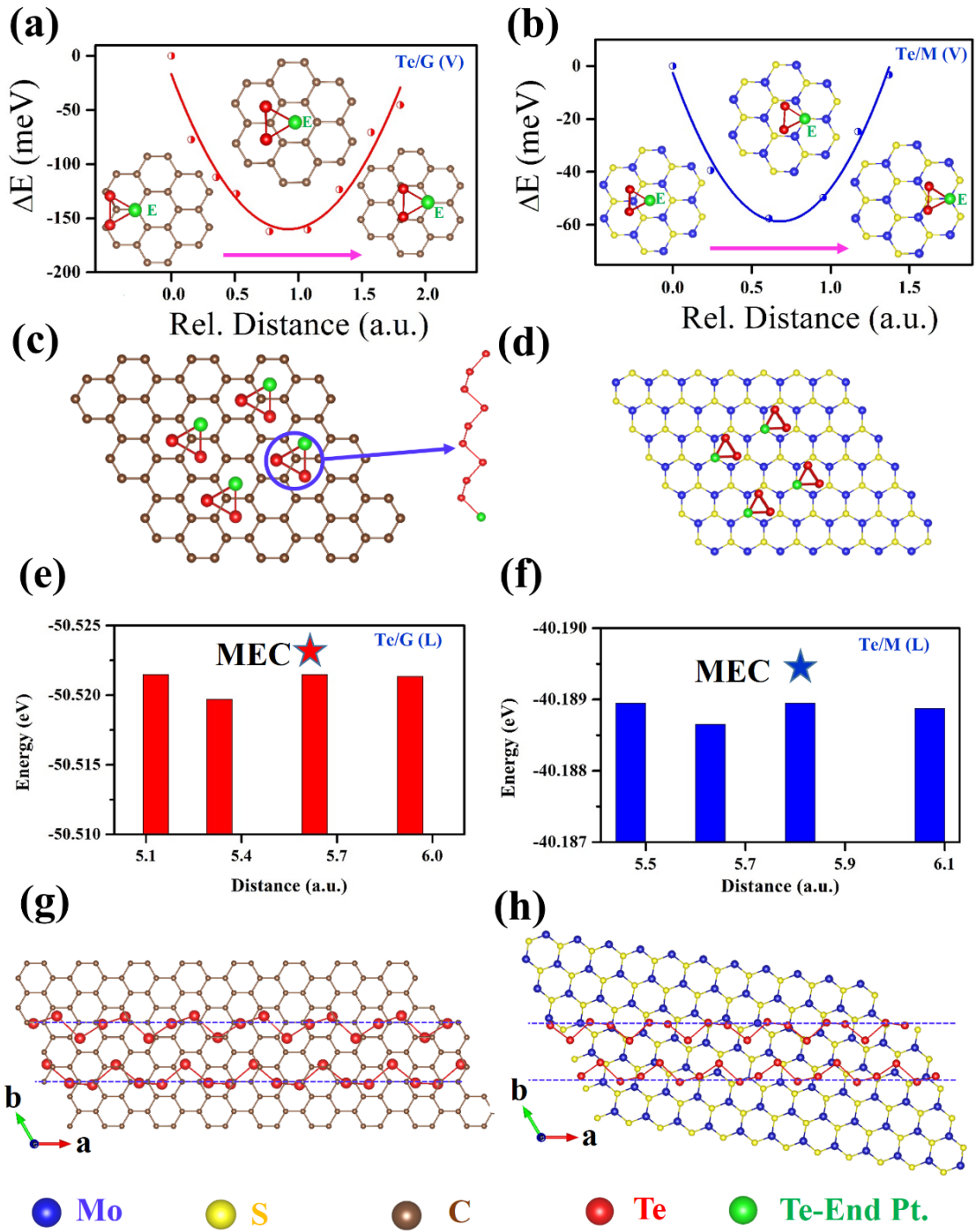

**Figure 1:** Ground state energies (with respect to a fixed vertex point) and corresponding structural images of a) Te/G (V) and b) Te/M (V) HJs. Lowest energy structure of c) Te/G (V), d) Te/M (V). Ground state energies (with respect to a fixed line, parallel to the NR axes and passing through the centre of the ML) of e) Te/G (L) and f) Te/M (L). The Minimum energy configurations (MEC) are indicated on each graph. The structural images corresponding to the minimum energy configurations of g) Te/G (L), h) Te/M (L).

The analysis implies that with respect to a single hexagon, the optimized position of the NR is at the center and at the vertex of the hexagons for graphene and MoS$_2$ respectively. In the next step, for higher densities of Te-NRs, Fig 1(c) and (d) present the top view of the energetically optimized vertical array of Te NR on the graphene and MoS$_2$ monolayer.

For lateral HJ, Fig 1(e) and (f) compares the formation energies of the HJ with the axial displacement of the two laterally placed Te-NR along a-axis with respect to a line parallel to the a-axis and passing through the center of the ML. For lateral HJs, the formation energies are very comparable with respect to such displacements. For two such NRs, the lateral hetero-structures are optimized as in Fig 1(g) and (h), corresponding to the minimum energy configuration (MIC) marked in Fig 1(e) and (f), after avoiding the electronic cross-talking from their periodic replication. An overall comparison of the formation energies per atom for the lateral and vertical hetero-structures indicates that formation of vertical HJ is more energetically favorable, as can be seen from Table S2 of the supplementary information (SI) [55]. Practically, the vertical growth of Te-NR may involve an epitaxial high-temperature process [56,57], whereas the synthesis of the lateral heterojunctions can be materialized by the mechanical processes like drop-casting or spin-coating of the 2D layers. After accomplishment of geometrical optimization, we proceed to investigate the electronic structure of all four possible heterojunctions (HJ).

1) Te/G(L) HJ: The elaborate electronic structure of this system is presented in Fig 2. During the formation of the HJ, with the Fermi-level($E_F$) of the array of Te - NR being ~ 2.5 eV above that of graphene, electrons are transferred from Te-NR to G, leading to an *n*-type doping[58]. The calculation of $E_F$ and the Bader analysis for charge transfer convey that per each Te atom ~ 1.2e are transferred from Te to G, as seen in Table 1. Intriguingly, this charge transfer is mostly located at the interface, as is evident from the charge density difference (CDD), calculated as $\rho_{CDD} = \rho_{1D/2D} - \rho_{1D} - \rho_{2D}$ with $\rho$ being the charge density, as plotted in Fig 2(c). This excess negative charge at the interface creates a space charge layer and thus hinders further electron transfer to graphene after creating a Schottky barrier. Such Schottky type HJ incites the opening of a direct band-gap of ~ 130 meV in graphene at the Γ-point, as is evident from the layer-projected band-structure in Fig 2(a). The schematic of the band-alignment diagram indicating the directional detail of carrier flow is presented in Fig 2(b). The corresponding density of states are presented at Fig S3 of SI. [55] The level projected charge densities for highest occupied (HOL), lowest unoccupied (LUL) and their adjacent significant levels are plotted in Fig 2(d), where the localized charges on Te-NRT at HOL-2 are transferred to the steeply dispersed levels of graphene at HOL, which are separated from the LUL and LUL+2 levels, contributed mostly due to graphene. The experimental photo-gating effects [1], as analyzed from the band-structure, reveals that the photo-generated holes can easily cross the barrier and thus the Te-side of the HJ accumulates the negative charge, leading to an illumination-induced control of the barrier-height. The magnitude of the barrier-height and the band-gap indicates the potential use of this system as a broadband detector in the MIR to NIR range.

2) Te/G(V) HJ: For the vertically stacked structure, the broken periodicity along the 2D plane and the resulting quantum confinement leads to highly localized energy levels of Te-NRs, almost un-hybridized with graphene bands, except at the Γ-point, as is evident from the Fig 2(e). The $E_F$ of the combined system shifts towards the conduction band, implying an *n*-type doping during the formation of the HJ. Bader analysis conveys a charge transfer of ~ 1.6e per atom of Te-NR. The corresponding band-alignment diagram after formation of HJ, shown in Fig 2(f), resemble the Z-scheme HJ, where its staggered nature, albeit similar to type II HJ, produces a different direction of the charge-flow. The proximity of valence band maxima (VBM) of Te-NR and conduction band minima (CBM) of graphene promotes a transfer of photo-generated electrons at the CB of graphene from the VB of Te-NR and thereby pertains a longer lifetime to the carriers due to an effective charge separation. The CDD and the level projected charge densities, as presented in Fig 2(g) and (h), conveys that the HOL and adjacent levels are contributed by Te-NR, whereas the LUL levels are having contributions from both graphene and Te-NR, as in corroboration with its band-structure. The system has a Γ-point direct band gap of ~ 150 meV and may have a potential application in MIR-NIR broadband detection as well as in photocatalytic energy harvesting. It also offers a possibility of formation of long-lived interlayer excitons with electrons and holes at the graphene and Te-NR layers due to the presence of Te-midgap levels, also having the contribution of structural reconstruction and defects[58,59].

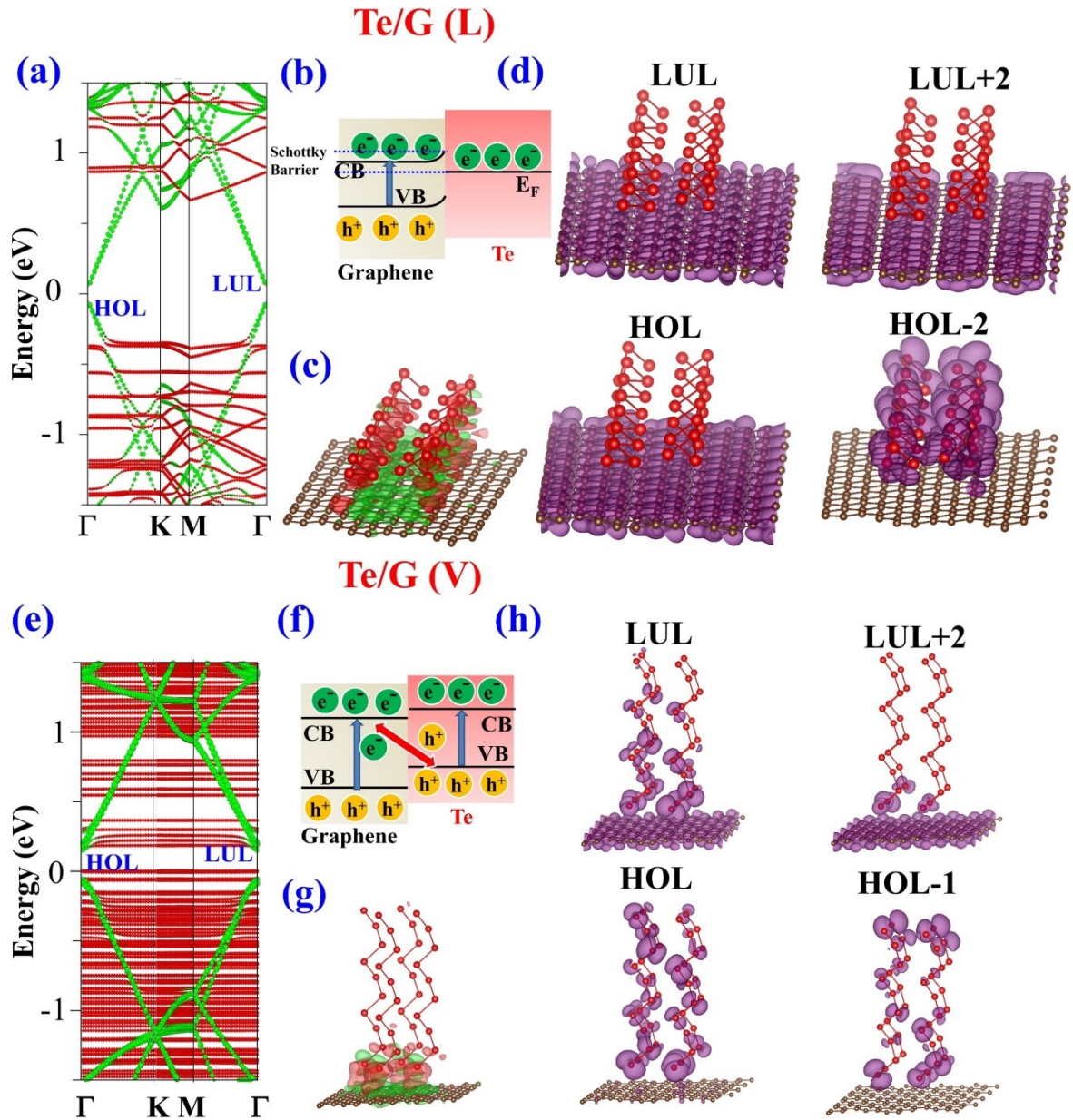

**Figure 2:** a) Band structure and b) corresponding band alignment of Te/G (L), c) charge density difference and d) band projected charge density of Te/G (L). e) Band structure and f) corresponding band alignment of Te/G (V), g) charge density difference and h) band projected charge density of Te/G (V).

3) Te/M(L) HJ: The band-structure of this HJ, as presented in Fig 3(a), clearly demonstrates a type-II HJ, having a staggered band-alignment as presented in Fig 3(b), where the VBM and CBM of the HJ are constituted of Te-NR and $MoS_2$ respectively, resulting an indirect band gap in the NIR range ~ 1.2 eV. The CDD, as presented in Fig 3(c) indicates the additional charge in the HJ to be mostly located at the interface. The LUL, HOL and their adjacent level projected charge densities, as presented in Fig 3(d), indicates them to be contributed by $MoS_2$ and Te-NR levels. During formation, there is a charge transfer of ~ 1.1e per atom of Te to the $MoS_2$, according to Table 1 of Bader analysis, implying an n-type doping. A type-II HJ is the most interesting among all the plausible types of HJ, having the potential for both broadband detection and energy-harvesting applications[60]. The band-alignment diagram in Fig 3(c) clearly indicates the potential of the present HJ in energy harvesting in the NIR and Vis range. In addition, it promotes the possibilities of formation of interlayer-excitons with a longer life-time. The photo-generated electron-hole pairs are located in two different layers, where, the conduction electrons of Te-NR are transferred to the CB of $MoS_2$ and the holes at the VB of $MoS_2$ are transferred to the VB of Te-NR. In this manner, the separation and migration of the carriers may pertain a longer lifetime to them, leading to a lesser rate of non-radiative recombination and thus resulting into a better performance of the HJ towards solar energy harvesting.

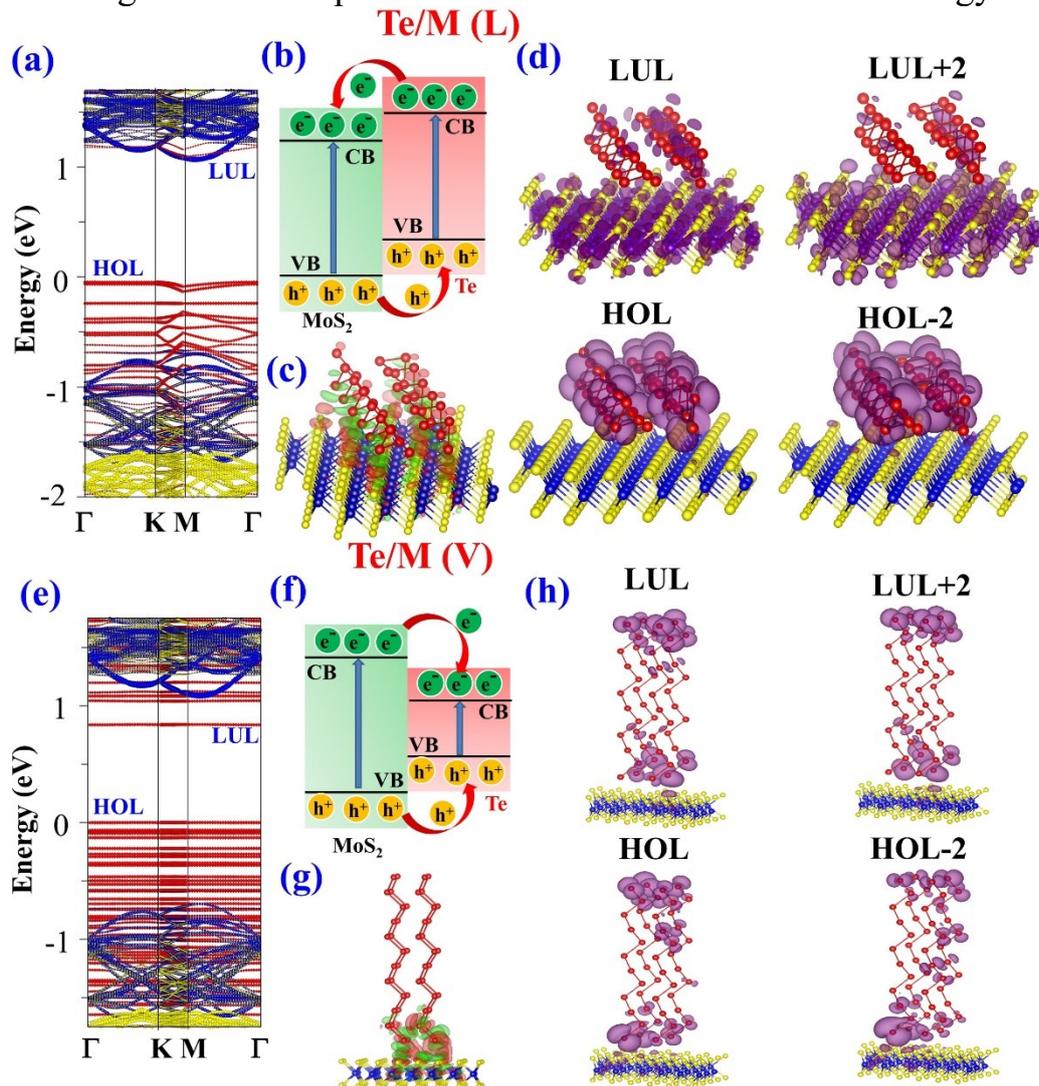

**Figure 3:** a) Band structure and b) corresponding band alignment of Te/M (L), c) charge density difference and d) band projected charge density of Te/M (L). e) Band structure and f) corresponding band alignment of Te/M (V), g) charge density difference and h) band projected charge density of Te/M (V).

4) Te/M(V) HJ: The level-projected band-structure of this HJ is presented in Fig 3(e), where, in a similar manner as the Te/G(V) system, the quantum confinement of Te-NR produces un-hybridized localized levels. There is a per atom charge transfer from Te-NR to $MoS_2$ of ~ 0.8e, expediting an n-type doping. The corresponding band-alignment diagram of the combined HJ, as depicted in Fig 3(f), indicates that both the VBM and CBM of the combined system are contributed from the Te-NRT and are located within the band-gap of $MoS_2$. Therefore, the holes and the electrons from the VB and CB of $MoS_2$ will populate the VB and CB of Te-NRT. Whereas the CDD is mostly confined near interface, as seen from Fig 3(g), the level-projected charge densities for HOL, LUL and their adjacent levels are entirely contributed by Te-NRT, at par with its band-structure. In this HJ, the localized Te levels at mid-gap may act as trap states to the photo-generated electrons and thus may render a longer lifetime of the excitons[58,59]. The HJ is having a direct band gap of 0.83 eV and can again be used as a broadband photodetector in the NIR range.

*Table 1:* *Bader Charge analysis and Fermi Energy Table of the different HJs.*

| System | | Te/G (V) | Te/G (L) | Te/M (V) | Te/M (L) |
|---|---|---|---|---|---|
| **Bader Charge (per atom)** | Te | -1.642 | -1.248 | -1.112 | -0.777 |
| | G | 0.006 | -0.017 | - | - |
| | $MoS_2$ | - | - | 0.858 | 0.172 |

The potential of optical application of these four HJs, as obtained from the combined electronic structure, can be further extended towards the calculation of their optical absorbance. The details of the calculation of the optical response are outlined in SI. [55] Fig 4(a) – (d) present the static optical absorbance per unit optical path for these four HJs, as calculated from the density functional theory (DFT) by using the real ($\varepsilon_1$) and imaginary ($\varepsilon_2$) parts of the dielectric response as related by the Kramer-Kronig relation, the implications of which can be summarized as follows. First, the peaks of the optical activities for Te/G(L) and Te/G(V) HJs are in the MIR range (2500 - 25000nm) and those for the Te/M(L) and Te/M(V) systems are in the NIR + Vis range (380 – 2500 nm). Second, the highly anisotropic nature of the hybrid systems is manifested in the difference of the in-plane and out-of-plane components of their respective absorbance. On the other hand, Fig 4(e) – (h) represents a direct comparison of the time-dependent optical absorbance of all four hybrid systems with respect to those of the pristine ones, computed using TDDFT and conveying the following observations. First, the flat optical response of graphene, as presented by the blue line in Fig 4(e) and (f) is modified into a broadband distribution from with its peak absorbance in the MIR range (2500 - 25000nm). Second, For $MoS_2$, the TDDFT spectra of both the pristine and hybrid structures are distinctively different in their inherent intricacies from the DFT ones, containing mid-gap states and clear signatures of

A and B excitons. Third, in the TDDFT spectra for the HJs of MoS$_2$, there is a significant shift in the energy positions of the A and B excitons toward higher range by ~ 1 eV, which points towards an enhanced lifetime of both the excitons. There is also a consequential increase in the binding energy of both A and B excitons of MoS$_2$ by ~ 200 and 300 meV for lateral and vertical HJs respectively. Henceforth, the range of optical activity of the HJ encompasses the MIR, NIR and the Vis range (from 380 – 2500 nm), as seen in Fig 4(g) and (h).

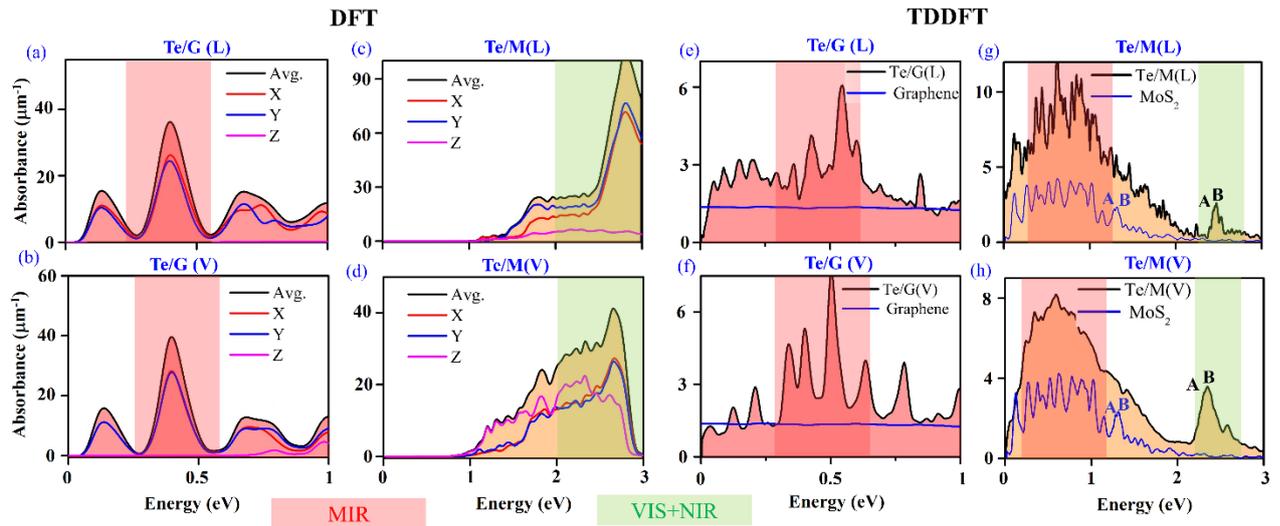

**Figure 4:** Static absorbances (from DFT) per unit optical length are plotted for a) Te/G(L), b) Te/G(V), c) Te/M(L) and d) Te/M(V) with their corresponding range of optical activity as shown in the lower panel. The TDDFT absorbance per unit optical length are plotted for e) Te/G(L), f) Te/G(V), g) Te/M(L) and h) Te/M(V) with their corresponding range of optical activity as shown in the lower panel.

It may also be mentioned in passing that all of these improved functionalities are highly sensitive of the optimized density of Te NRs on top of the 2D ML. With uncontrolled density, the resulting highly hybridized interaction is detrimental to the semiconducting properties, as will be evident from Fig S4. [54]

Formation of hybrid HJ also incorporates additional functionalities over the pristine systems. The enhanced DFT+SOC non-collinear spin-densities for all the hybrid HJs and the corresponding spin-projected DOS are plotted in Fig S5, S6 and S7 respectively. [54] The interaction of the hybrid systems and its feeble spin-polarization with the plane of polarization of the applied illumination also gives rise to several optical phenomena like linear dichroism, magneto-optic and electro-optic Kerr and Faraday effects. The difference between the absorption of the light polarized perpendicular and parallel to the out-of-plane axis can be computed from the linear dichroism. Interaction of the polarized spins of the system with the applied magnetic and electric fields of the incident light results into a rotation of its plane of polarization. These magneto-optical and electro-optical effects are amplified by at least an order of magnitude after formation of hybrid HJs, as can be seen from the Fig S8 [54].

In conclusion, by using first-principles calculations, we predict an effective design of a 1D/2D hybrid HJ, by using a single helix of Tellurene and the monolayers of 2D materials like graphene and $MoS_2$. The lowest energy configurations of the four hybrids manifest the type-I, type-II, Z-scheme and Schottky-type heterojunctions, having excellent potentials for application in the broad-band photo-detection, energy harvesting and longer quasiparticle lifetime. The broad range of optical activities for these HJs encompasses the MIR to NIR and Vis regions and thus predicts excellent application potential of these systems in the field of photonic IR detectors and green energy materials.

**Acknowledgement:**
DK acknowledges funding from the Knut and Wallenberg foundation and computational resources from SNIC. DK also acknowledges discussion with Weine Ollovsson regarding the resources of TDDFT runs. TKM and DK sincerely thanks Arindam Ghosh for several fruitful discussions.

# Supporting Information:
# 1D/2D hybrid Te/Graphene and Te/MoS$_2$: multifaceted broadband photonics and green energy applications

Tuhin Kumar Maji[1], Kumar Vaibhav[2], Anna Delin[3], Olle Eriksson[4, 5]\*, Debjani Karmakar[4,6]\*

[1] Department of Physics, Indian Institute of Science Bangalore, Bangalore, 560012, India
[2] Computer Division, Bhabha Atomic Research Centre, Trombay, Mumbai 400085, India
[3] Swedish e-Science Research Center (SeRC), KTH Royal Institute of Technology, SE-10044 Stockholm, Sweden
[4] Department of Physics and Astronomy, Uppsala University, Box 516, SE-75120 Uppsala, Sweden
[5] School of Science and Technology, Örebro University, Fakultetsgatan 1, SE-70281 Örebro, Sweden
[6] Technical Physics Division, Bhabha Atomic Research Centre, Trombay, Mumbai 400085, India

\*Corresponding Author: debjan@barc.gov.in/olle.eriksson@physics.uu.se

**Computational Methodology:**

To study the electronic properties of the Te/M interface or Te/G interface, we have initially constructed interfaces by joining the two lattices by the coincidence site lattice (CSL) method, as implemented in the Atomistic Toolkit 15.1 package [1]. Detail of the methodology is described in the Ref. 13 of the main manuscript. A 4×3×1 Graphene or $MoS_2$ layer has been taken as the base 2D structure and Te rods are grown vertically or horizontally on the top of it. In all the cases the strin has been minimized.

For the static density functional calculation, we have used spin-polarized plane-wave pseudopotential methods with projector augmented wave (PAW) formalisms as implemented in the Vienna Ab initio Simulation Package (VASP) [2,3]. The ionic positions of the individual interfaces are relaxed to obtain the ground state energy configuration. A vacuum of ~10 Å is introduced at the top and bottom of the interface to avoid periodic replication of the system. The exchange correlation interactions are treated with the generalized gradient approximation (GGA) with Perdew-Burke-Ernzerhof (PBE) functionals [4]. Interface-induced van der Waals interactions are taken care of after including a semiempirical dispersion potential to the density functional theory (DFT) energy functional according to the Grimme DFT-D2 method [5]. The cutoff energy for the plane-wave expansion is set as 500 eV and a Monkhorst-Pack grid of 5×5×3 is used for Brillouin zone sampling for all calculations. The ionic positions and the lattice parameters are relaxed by using the conjugate gradient algorithm until the Hellmann-Feynman force on each ion is less than 0.01 eV.

The static optical properties and the DFT quantum transport calculations are performed using the Atomistic Toolkit 15.1 package package[1], with the GGA-PBE variant of exchange correlation.

The time dependent properties are calculated with the all electron full-potential linearized augmented plane wave approach including local orbitals (FP-LAPW + lo) as implemented in the ELK7.4 code. The exchange-correlation potentials are calculated with local density approximation (LDA) with Perdew-Wang / Ceperley Alder functional. The interstitial plane wave vector cut-off $K_{max}$ is chosen such that $R_{mt}K_{max}$ equals 7 for all the calculations, with $R_{mt}$ being the smallest of all atomic sphere radii. The convergence criteria for total energy and RMS change in Kohn-Sham potential are kept as 0.0001 and 0.000001 eV respectively. The valence wave functions inside the spheres are expanded upto $l_{max}$ =10 and the charge density was Fourier expanded upto $G_{max}$ = 12.

# Optical Property Calculation

## A. TDDFT Formulation

In this section, we will briefly describe the TDDFT methodology used in this present study. TDDFT aims to map the time-dependent SchrÖdinger equation onto an effective one-electron problem. Here, time dependence is incorporated in the approximation of the exchange-correlation kernel (xck) from the explicit time dependence of the exchange-correlation potential and electron-density as:

$$f_{xc}(r, r', t - t') = \frac{\partial v_{xc}(r,t)}{\partial \rho(r',t')} \tag{1}$$

Optical absorption calculation includes many-body purturbative method of solution of Bethe Saltpeter Equation (BSE) using the one-body Green's function. In BSE framework, we can write the dielectric function in terms of the xck as:

$$\epsilon^{-1}_{GG'}(q, \omega) = \delta_{GG'} + v_{GG'}(q)\chi'(q, \omega), \tag{2}$$

where $v(q)$ is the bare Coulomb potential and $\chi$ is the full response function. $\chi$ is associated to the response function $\chi^0$ of the non-interacting Kohn-Sham system as:

$$\chi'(q, \omega) = \frac{v_{GG'}(q)\chi^0_{GG'}(q,\omega)}{1-[v_{GG'}(q)+f_{xc}(q,\omega)]\chi^0_{GG'}(q,\omega)} \tag{3}$$

One can have a rigorous solution of BSE. However. the frequency independent approximation for the xck, which is known as the Bootstrap kernel commonly used to solve BSE because it is computationally inexpensive. We can write BSE as:

$$f_{xc}^{TDDFT}(q, \omega) = \frac{\epsilon^{-1}(q,\omega=0)}{\chi^0_{GG'}(q,\omega=0)} \tag{4}$$

The coupled equations (2), (3) and (4) are solved by initially setting $f_{xc}^{TDDFT} = 0$ and then calculating $\chi'(q, \omega)$ and thus $\epsilon^{-1}_{GG'}(q, \omega)$. This value is then employed to equation (6) to get the new $f_{xc}^{TDDFT}$. This process is repeated until the self-consistency is obtained at ω = 0.

The long-range component of exchange-correlation kernel is frequency independent having the form -$\alpha^{static}$ / $q^2$. However, more accurate formulation includes frequency dependence. The long-range contribution kernel (LRC) integrates the frequency dependence in the following form into the dynamical exchange:

$$f_{xc}^{dyn}(q, \omega) = -\frac{1}{q^2}(\alpha + \beta\omega^2) \tag{5}$$

The inclusion of frequency in dynamical kernel has improved the calculations to acquire important information on the exciton as well as charged excitons[2].

### B. DFT optical properties

In a dipole approximation, the real and imaginary parts of the interband dielectric response function in tensor form can be calculated in terms of the momentum-matrix element between the occupied and unoccupied levels, using Fermi-golden rule as:

$$\epsilon^1_{\alpha\beta}(\hat{q},\omega) = 1 + \frac{4\pi e^2}{V} \lim_{q \to 0} \frac{1}{q^2} \sum_{nm\,k} 2f_{nk} \langle u_{mk+e_{\alpha}q} | u_{nk} \rangle \langle u_{nk} | u_{mk+e_{\beta}q} \rangle \times \left( \frac{1}{\epsilon_{mk}-\epsilon_{nk}-\omega} + \frac{1}{\epsilon_{mk}-\epsilon_{nk}+\omega} \right) \tag{6}$$

and

$$\epsilon^2_{\alpha\beta}(\hat{q},\omega) = \frac{4\pi^2 e^2}{V} \lim_{q \to 0} \frac{1}{q^2} \sum_{nm\,k} 2f_{nk} \langle u_{mk+e_{\alpha}q} | u_{nk} \rangle \langle u_{nk} | u_{mk+e_{\beta}q} \rangle \times (\delta(\epsilon_{mk}-\epsilon_{nk}-\omega) - \delta(\epsilon_{mk}-\epsilon_{nk}+\omega)) \tag{7}$$

Thus, the absorbance can be calculated from:

$$\sigma(\omega) = \sqrt{2}\omega \sqrt{\frac{\varepsilon_2(\omega)}{\varepsilon_1(\omega)} + \varepsilon_2^2 - \varepsilon_1(\omega)^{1/2}} \tag{8}$$

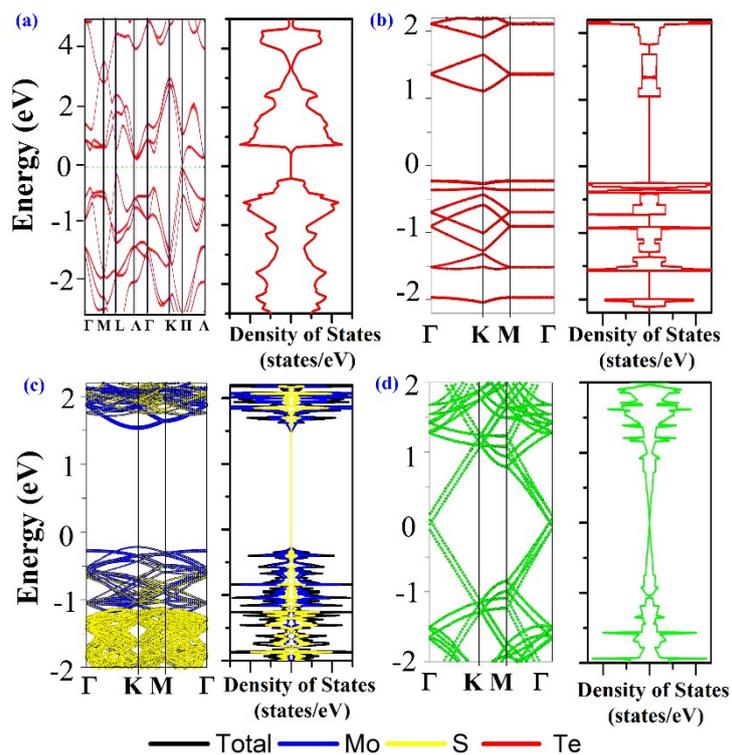
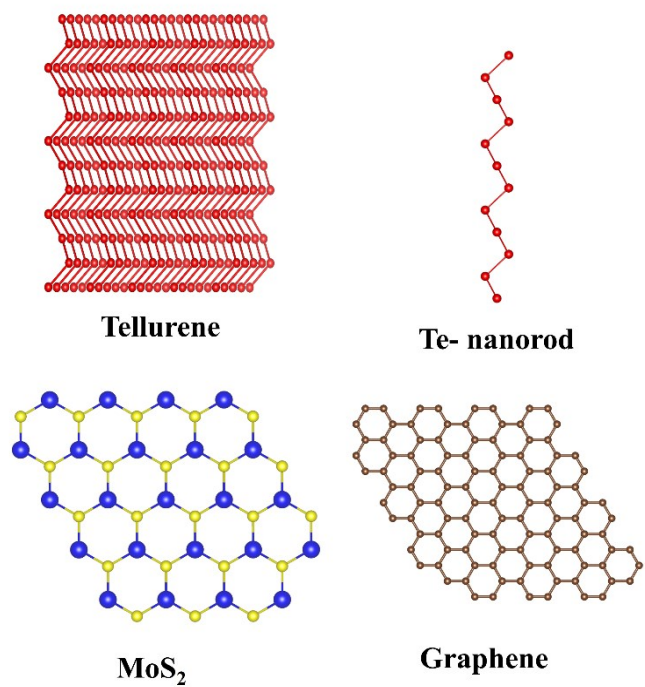

**Figure S1:** Band structure and corresponding density of states of a) Te-bulk, b)Te-nanorod c) MoS$_2$ and d) Graphene. The corresponding system figures are presented in the lower panel.

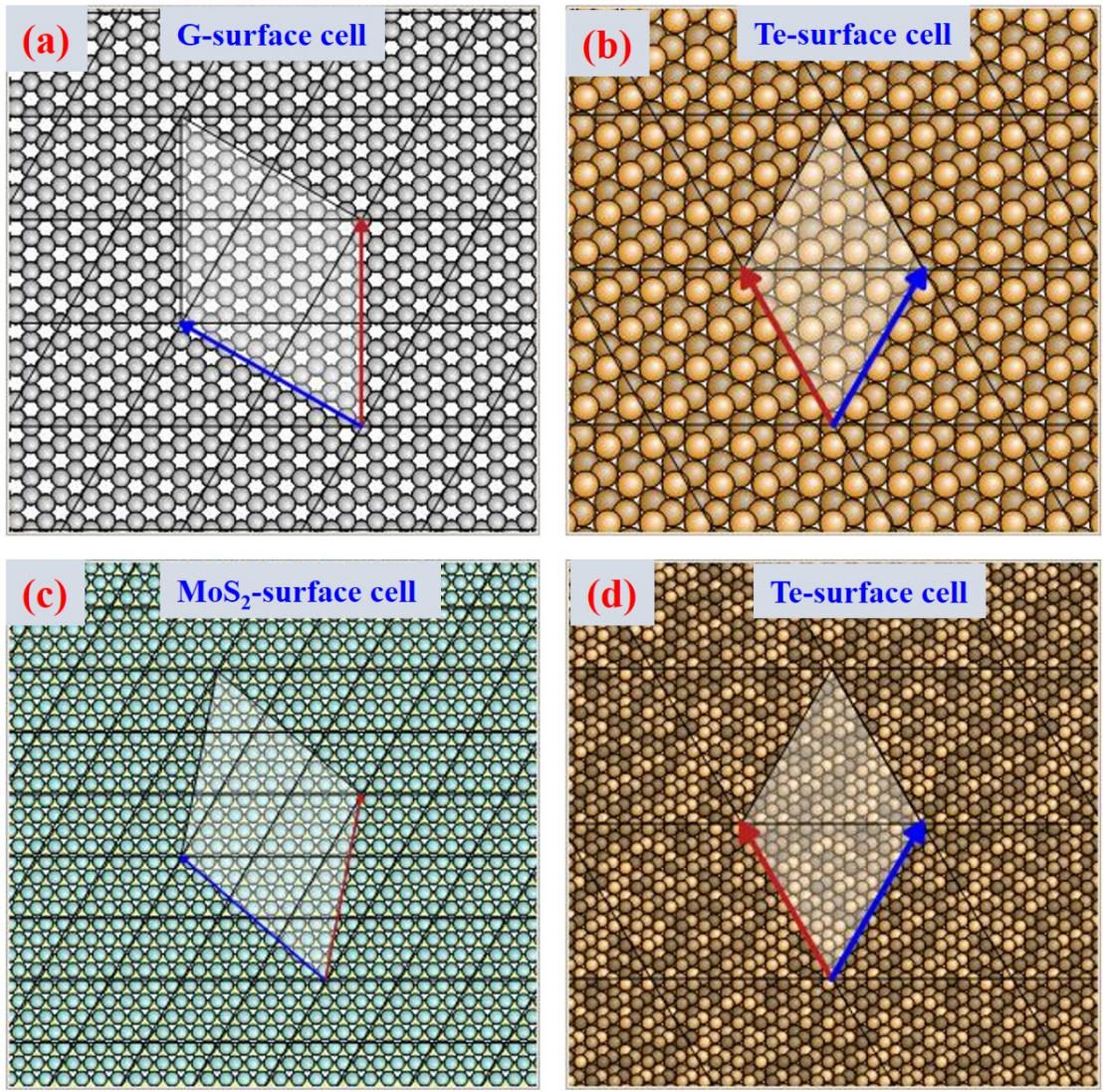

**Figure S2:** Construction of vertical heterostructure a) Graphane surface cell, b) Te surface cell of Te/G (V). Construction of vertical heterostructure c) MoS$_2$ surface cell, d) Te surface cell of Te/M (V).

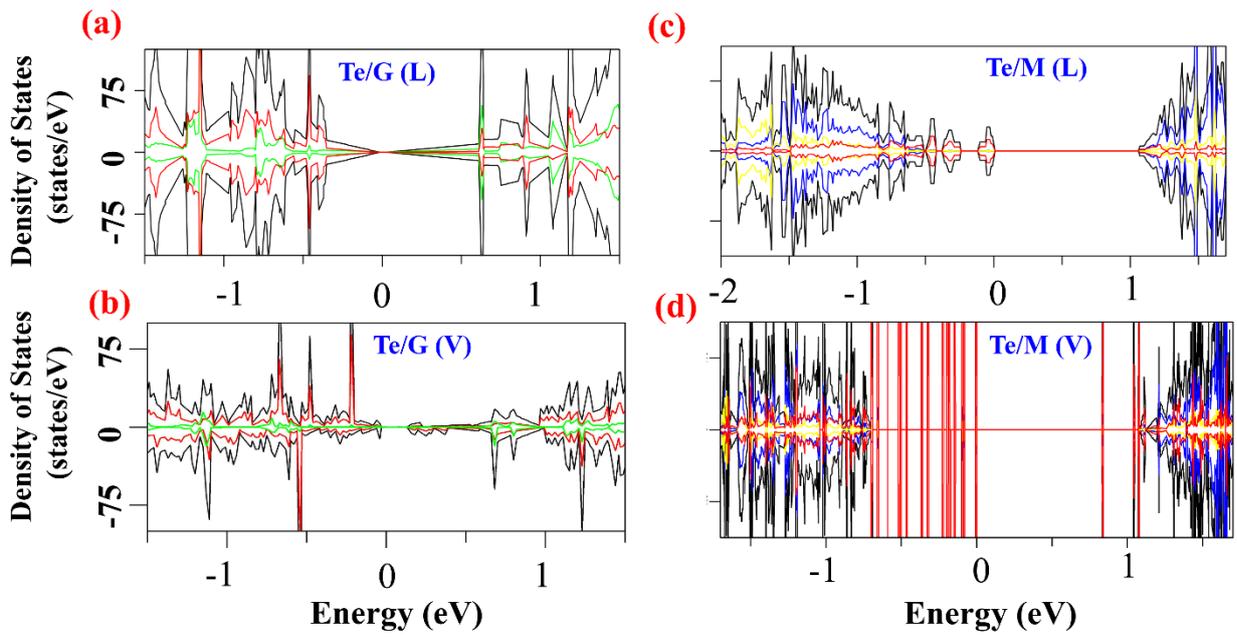

**Figure S3:** Density of states of a) Te/G (L) b)Te/G (V) c) Te/M(L) and d) Te/M (V).

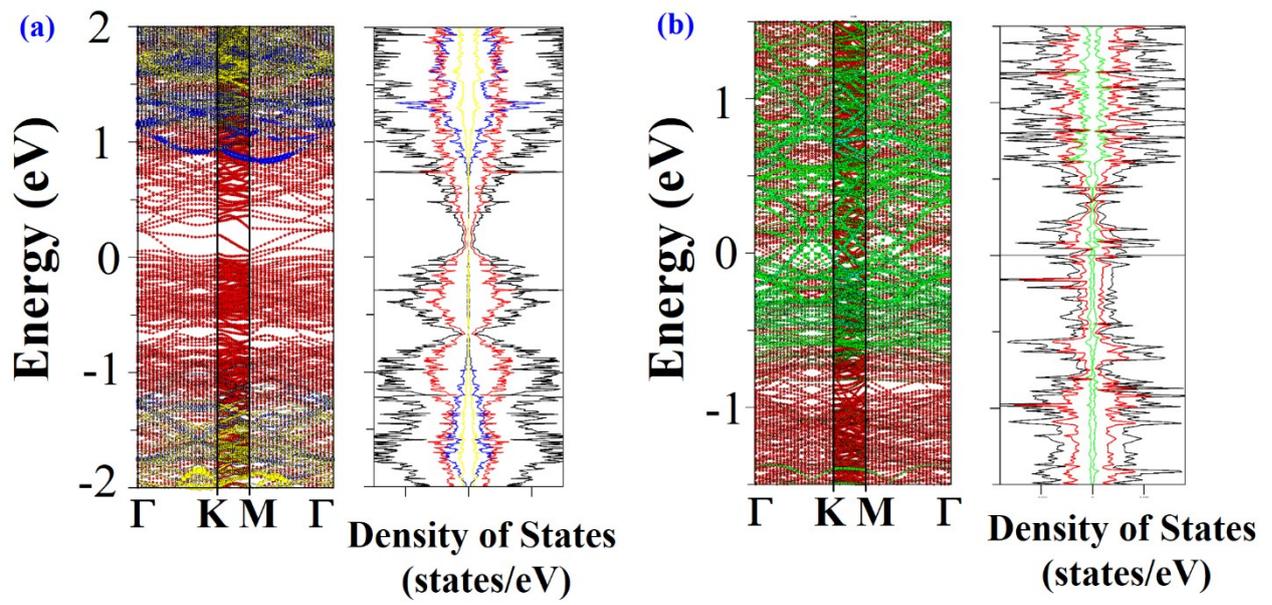

**Figure S4:** Band structure and corresponding density of states (DOS) of a) Te/M (V) and b) Te/G (V) for higher density of vertical Te NRs.

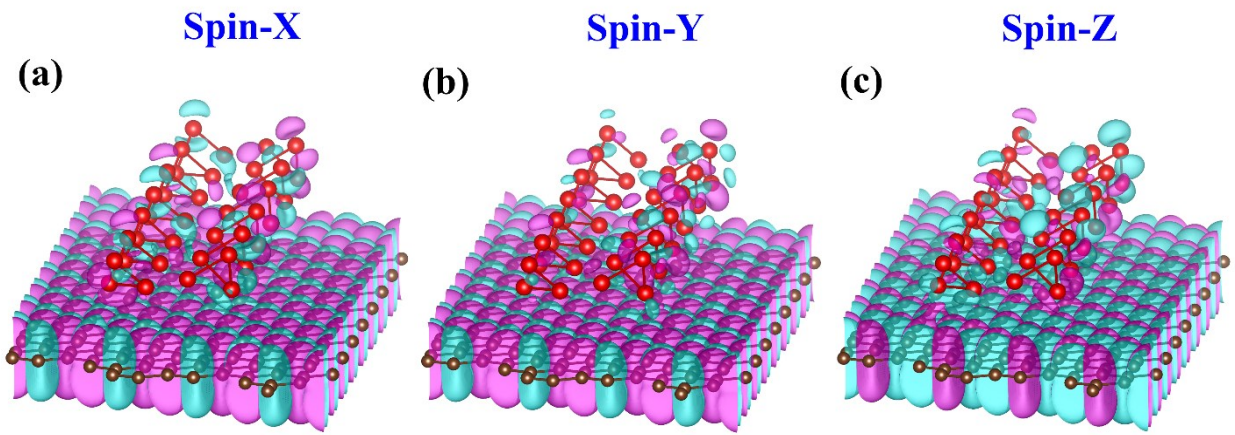

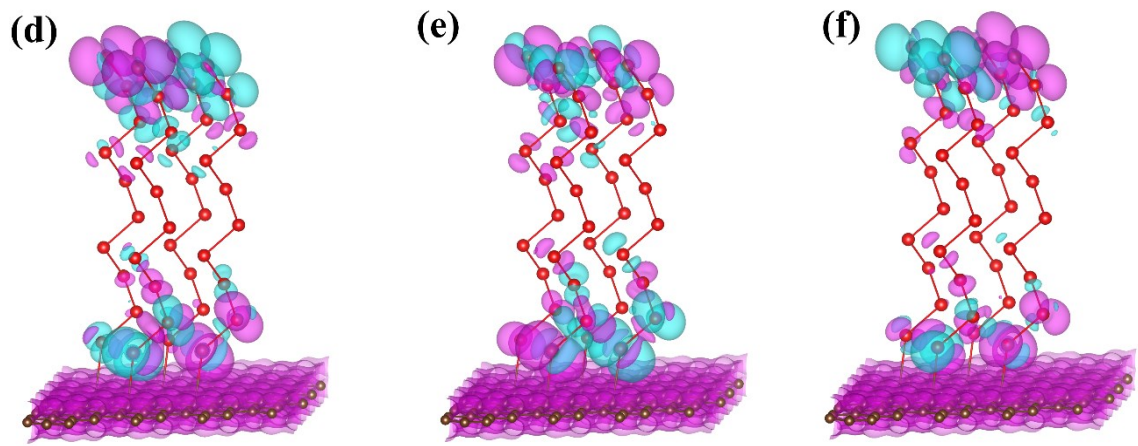

**Figure S5:** a) X, b) Y, c) Z directional spin density of Te/G(L); e) X, f) Y, g) Z directional spin density of Te/G(V)

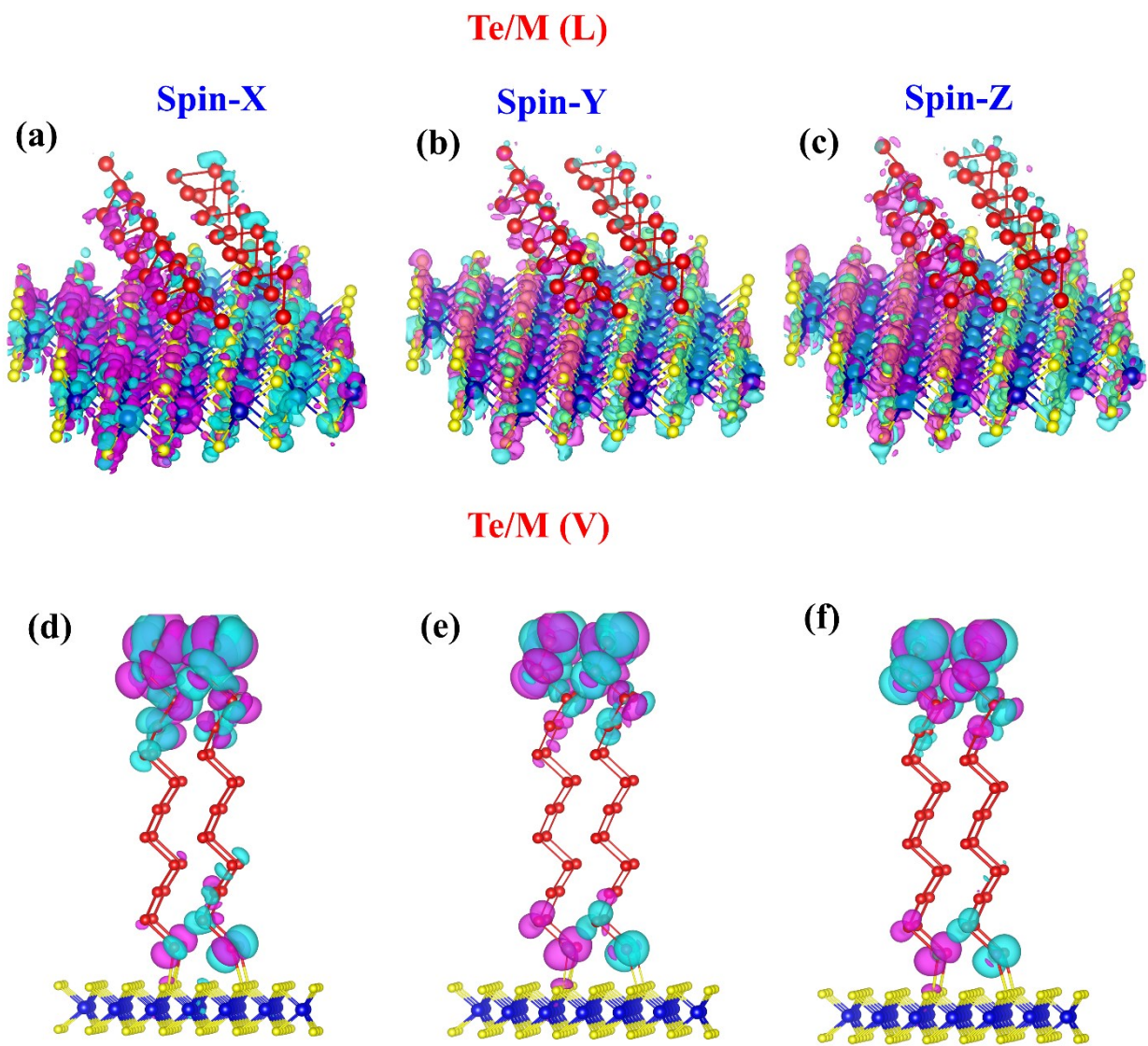

**Figure S6:** a) X, b) Y, c) Z directional spin density of Te/M(L); e) X, f) Y, g) Z directional spin density of Te/M(V).

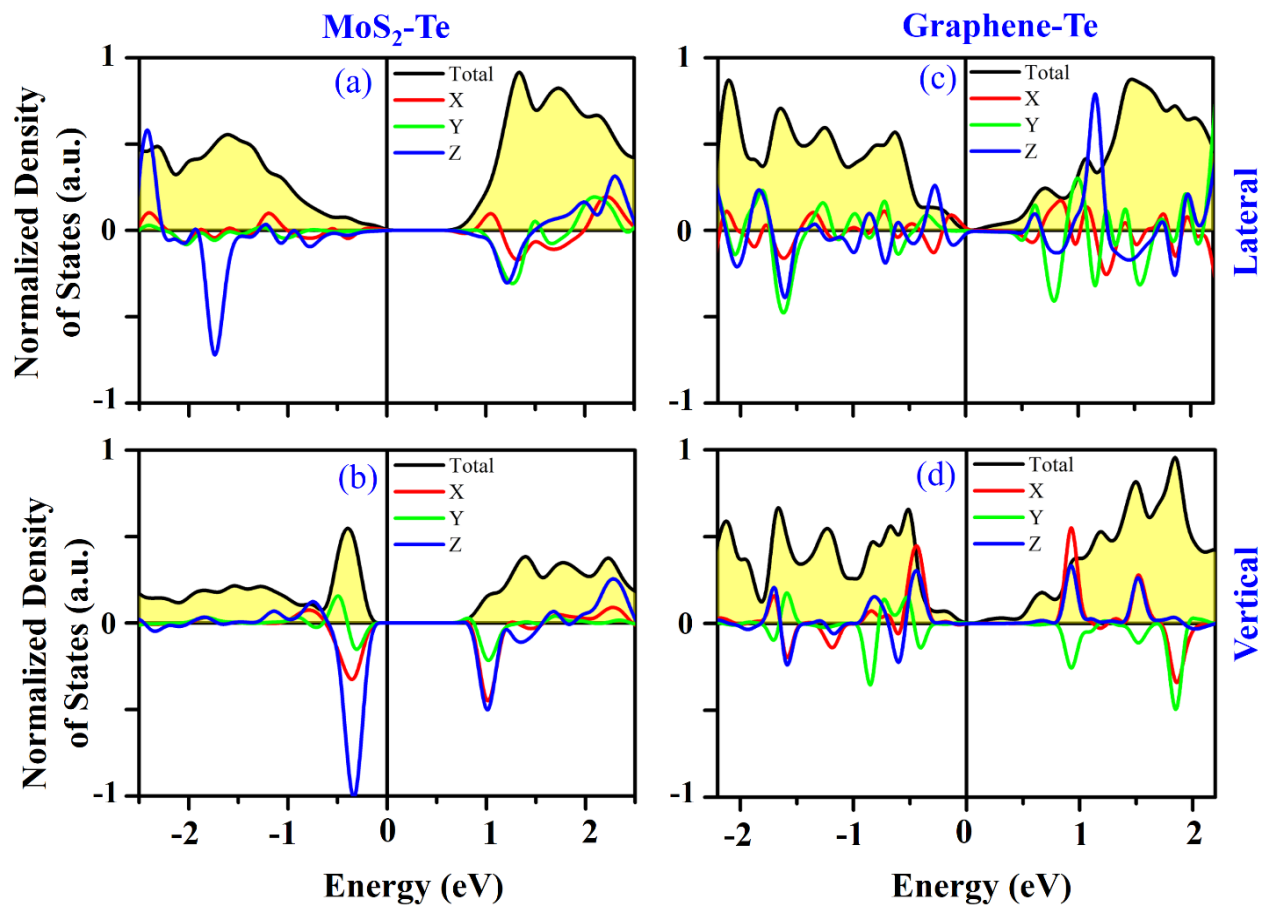

**Figure S7:** Spin-orbitally coupled density of states (SODOS) plot of a) Te/G (L), b) Te/G (V), c) Te/M(L) and d) Te/M(V).

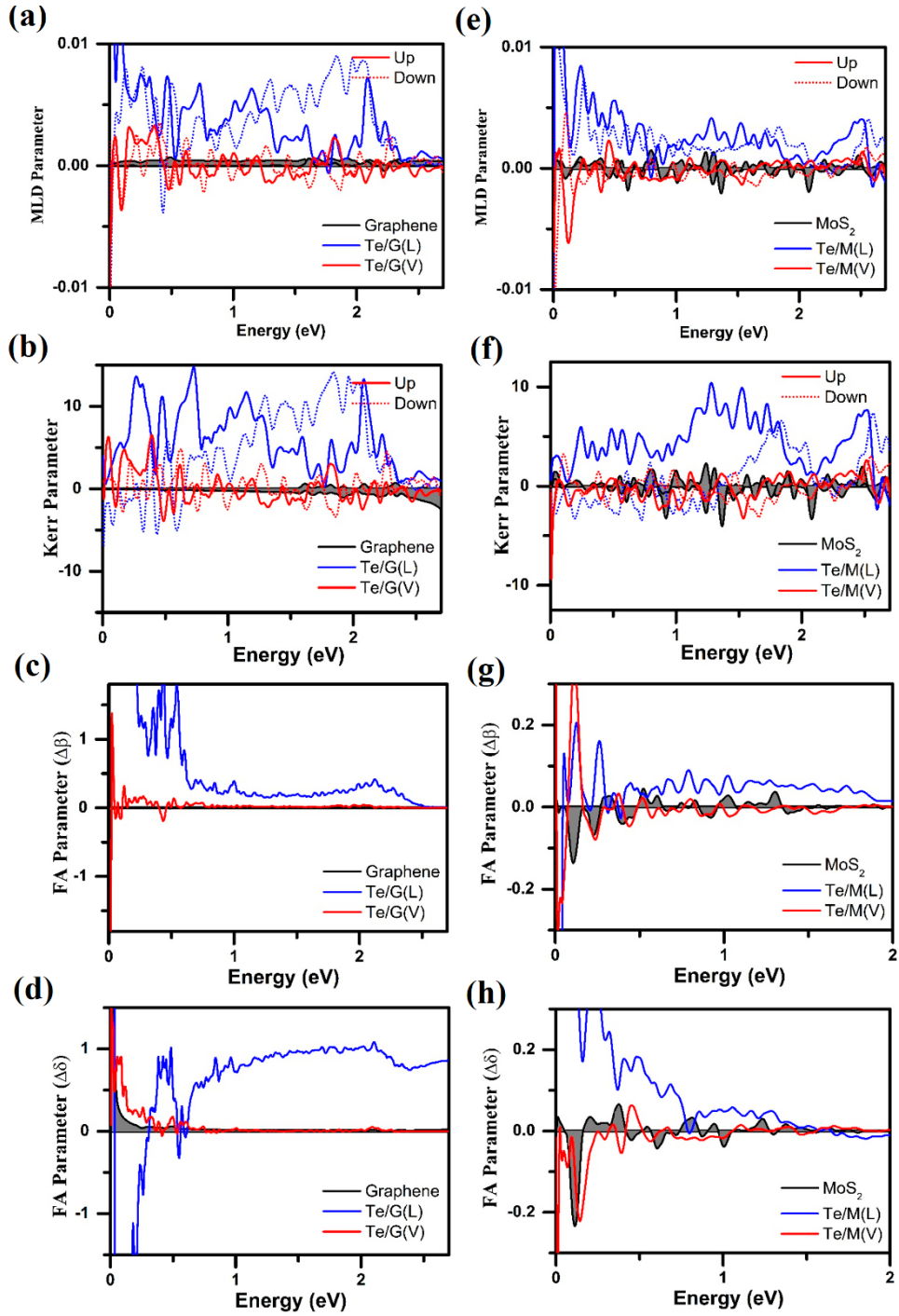

**Figure S8:** Comparison of (a) Linear dichroisom, (b) Kerr rotational angle, (c) Faraday rotational angle for interaction with the magnetic and (d) Faraday rotational angle for interaction with the electric field for G and the corresponding HJ; similar comparison of (e) Linear dichroisom, (f) Kerr rotational angle, (g) Faraday rotational angle for interaction with the magnetic and (h) Faraday rotational angle for interaction with the electric field for M and the corresponding HJs.

**Table S1:** Magnetic Moment table of different systems

| System | Te/G (V) | Te/G (L) | Te/M (V) | Te/M (L) | Graphene | MoS$_2$ |
|---|---|---|---|---|---|---|
| Mag.Mom. (μB) | 0.225 | 0.545 | 0.0021 | 0.0002 | 0.0003 | 0.000 |

**Table S2:** Formation energy of different systems

| System | Te/G (V) | Te/G (L) | Te/M (V) | Te/M (L) |
|---|---|---|---|---|
| FE (/ atom) (eV) | -0.0744 | -0.026 | -0.261 | -0.033 |

**Table S3:** Match properties of different parameters for vertical systems

| Parameters | Te/G (V) | Te/G (M) |
|---|---|---|
| No of atom | 288 | 1104 |
| Area | 251.81 Å$^2$ | 968 Å$^2$ |
| Angle between vectors | 60° | 60° |
| Angle between Surfaces | 150° | 160.89° |
| Mean Abs. strain | 2.76% | 3.99% |

**Table S4:** Excitonic position of MoS$_2$ before and afte Te-nanorod attachment

|  | A-exciton (eV) | B-exciton (eV) | Band Gap (eV) | A-exciton BE (eV) | B-exciton BE (eV) |
|---|---|---|---|---|---|
| MoS$_2$-Te (Lat) | 2.43 | 2.47 | 1.76 | 0.67 | 0.71 |
| MoS$_2$-Te (Ver) | 2.31 | 2.34 | 1.75 | 0.56 | 0.59 |